\documentclass[%
 reprint,
 amsmath,amssymb,
 aps,
]{revtex4-2}
\usepackage{physics}
\usepackage{amsmath}
\usepackage{amssymb}
\usepackage{graphicx}
\usepackage{dcolumn}
\usepackage{bm}
\usepackage[version=4]{mhchem}

\begin{document}

\preprint{APS/123-QED}

\title{High-throughput optical absorption spectra for inorganic semiconductors}

\author{Ruo Xi Yang}
\email{yangroxie@gmail.com}
\affiliation{Materials Science Division, Lawrence Berkeley National Lab}

\author{Matthew K. Horton}
\affiliation{Materials Science Division, Lawrence Berkeley National Lab}

\author{Jason Munro} 
\affiliation{Materials Science Division, Lawrence Berkeley National Lab}

\author{Kristin A. Persson}%
\affiliation{Molecular Foundry, Lawrence Berkeley National Lab}%
\affiliation{Department of Materials, University of California, Berkeley}

\begin{abstract}
An optical absorption spectrum constitutes one of the most fundamental material characteristics, with relevant applications ranging from material identification to energy harvesting and optoelectronics. 
However, the database of both experimental and computational spectra are currently lacking. 
In this study, we designed a computational workflow for optical absorption spectrum and integrated the simulated spectra into the Materials Project.
Using density-functional theory, we computed the frequency dependent dielectric function and the corresponding absorption coefficient for more than 1000 solid compounds of varying crystal structure and chemistry. 
The computed spectra show excellent agreement, as quantified by a high value of the Pearson correlation, with experimental results when applying the band gap correction from the HSE functional.  The demonstrated calculated accuracy in the spectra suggests that the workflow can be applied in screening studies for materials with specific optical properties. 

\end{abstract}

\maketitle


\section{Introduction}
Optical absorption spectroscopy is an important material characterization for identifying material species, atomic structures, reactions, and for understanding the electronic structure and light-matter interaction.\cite{kaufmann_ultraviolet_2002, zhang_characterization_2019} 
Optical light wavelengths, spanning UV, visible light and near-IR (350 - 750 nm), are used to probe materials including crystalline semiconductors, molecules, metals and amorphous materials.   
Due to the direct relation between structure and physical properties, \textit{in situ} UV-vis spectroscopy is commonly used to monitor the crystallization, composition, and phase transition at relevant time and length scales by tracking the evolving optical properties.\cite{zhang_situ_2019, buckner_situ_2019, babbe_optical_2020} 
Its non-destructive nature and inexpensive access makes it a popular tool to probe dynamic excitation processes in materials. 

The absorption process entails the interaction between incident electromagnetic waves and the material electronic structure.
Specifically in semiconductors, the electrons are excited from the valence band to the conduction band by absorbing the photon energy. Therefore, optical spectroscopy is often used to determine the optical bandgap ($E_\mathrm{g,opt}$) by considering the onset of the absorption edge.\cite{dolgonos_direct_2016}
Intrinsically, the optical bandgap is largely determined by the electronic bandgap  $E_\mathrm{g,e}$ between the band edges, considering only vertical transitions due to lack of momentum transfer from the photons ($q \rightarrow 0$). 
However, in some materials where direct forbidden transitions are present, $E_\mathrm{g,opt}$ can be larger than $E_\mathrm{g,e}$.
Additionally, when electron-hole Coulomb interaction is strong, resonant peaks of exitons form below the band-to-band absorption edge, leading to a below-bandgap $E_\mathrm{g,opt}$. \cite{koch_semiconductor_2006,rohlfing_excitonic_1998-1, elliott_intensity_1957, ogawa_optical_1991}. 
Moreover, charged defects with energy levels close to the band edges can cause the absorption edge to broaden and extend into lower energy range. \cite{redfield_effect_1963}
Extrinsically, the measured $E_\mathrm{g,opt}$ is influenced by many factors, including experimental setup, sample thickness etc.  
Various mathematical ways of extrapolating a linear fitting of the spectral tailing often result in inconsistent $E_\mathrm{g,opt}$ even in the same sample. \cite{raciti_optical_2017, zanatta_revisiting_2019}

These factors interplay and complicate the determination of E$_\mathrm{g,opt}$.
To this end, a database of internally consistent, computed optical absorption spectra using a standardized approach will be of benefit in providing a common reference point for experiments. 
By comparing with the absorption spectra of crystalline, defect-free materials, experiments can infer whether other influences, beyond band-to-band transitions, such as experimental setup and interpretation, defect formation, excitons, electron-phonon coupling etc., are present. 
From materials design perspective, these data provide an important yet currently not widely available material descriptor, the optical band gap $E_\mathrm{g,opt}$, for optoelectronic materials design.

In this study, we developed a high-throughput computational workflow for generating optical absorption spectrum using density functional theory (DFT) as implemented in the VASP package,\cite{kresse_efficient_1996-2, kresse_efficiency_1996-2}
with projector augmented wave core potentials (reciprocal space projection). \cite{blochl_projector_1994-2}
We used the independent-particle approximation (IPA) to compute the frequency dependent dielectrics for more than 1000 solid materials chosen from the Materials Project.\cite{jain_commentary_2013} 
Benchmarks against the random-phase approximations (RPA) and with experiments, using the Pearson correlation coefficients as metric, show excellent agreement when incorporating a scissor shift of the difference between PBE band gap and HSE band gap. We anticipate that the workflow and growing dataset will be useful to the general materials community, and in particular for designing optoelectronic materials. 

\section{Methods}
 
For each material, the structure was relaxed using the PBEsol functional,\cite{perdew_restoring_2008-4,perdew_erratum_2009-2} followed by a more rigorous relaxation with the r2SCAN functional,\cite{furness_accurate_2020} as implemented in VASP. \cite{kresse_efficient_1996-2, kresse_efficiency_1996-2}.
The structures were relaxed until the forces were converged to  1 meV/\AA, and the energy to 10$^{-6}$ eV. 

The static DFT calculation was performed using the tetrahedron smearing method, with the final wavefunctions then used for the following optical calculation.

The optical calculation was performed within the independent-particle approximation (IPA), with two times the number of bands initially considered in the static calculation to ensure the absorption spectrum covers a sufficient energy range. 
The \textit{k}-point density is chosen to be high (400/atom) to ensure convergence against the density of states. 

Here, the imaginary part of the frequency dependent microscopic dielectric function is calculated :
\begin{equation}
\begin{split}
\epsilon_2 = \frac{4\pi e^2}{\Omega} \lim_{q\rightarrow0}\frac{1}{q^2} 
& \sum_{v,c,\mathrm{k}} 2 w_\mathrm{k} \delta(E_{ck} - E_{vk} - h\omega) \\ 
& \times \braket{u_{c,k + qe_\alpha}}{u_{vk}} \braket{u_{vk}}{u_{c,k+qe_\beta}}  
\end{split}
\end{equation}
where $\Omega$ is the unit cell volume, $e$ is the electron charge, $q$ is the photon momentum, $\omega$ is the photon frequency, and $c$, $v$ and $k$ denote the conduction band, valence band, and the k-point, respectively. $u$ is the electron wavefunction and the subscript denotes a k-point and band. Here, $\ket{u_{nk+qe_\alpha}}$ can be obtained using first-order perturbation theory. 

The real part of the dielectric function is calculated via the Kramers-Kronig relation:
\[ \epsilon_1(\omega) = 1 +\frac{2}{\pi} \int_0^{\infty}\frac{\epsilon_2(\omega')\omega'}{\omega'^2 - \omega^2} d\omega'\] 

The optical absorption coefficient relies on the extinction coefficient $\Tilde{k}$ and the refractive index $\Tilde{n}$, where 
\[ \Tilde{k}^2 = \Tilde{n}^2 - \epsilon_1 \]
and $\Tilde{n}$ follows:
\[ \Tilde{n} = \frac{\epsilon_2}{2\Tilde{k}} \]
The frequency-dependent absorption coefficient depends on the extinction coefficient and the incident photon frequency: 
\[ \alpha = \frac{2\omega}{c} \Tilde{k} \]
Combining the above equations, we arrive at the absorption coefficient $\alpha$ as a function of both the real and imaginary parts of the dielectric function:
\[  \alpha = \frac{2\pi E}{hc} \sqrt{2}\times \sqrt{ \sqrt{\epsilon_1^2 + \epsilon_2^2} - \epsilon_1}  \]
where $E$ is the incident photon energy and $c$ is the speed of light in vacuum. 

At the long-wavelength limit ($q \rightarrow 0$) the dielectric matrix determines the optical properties accessible to optical probes. 
In IPA, the macroscopic dielectric is approximated to be the same as the microscopic dielectric by neglecting the local field effect, hence $\epsilon_{\mathrm{mac}} \approx \mathrm{lim}_{\mathbf{q} \rightarrow 0} \epsilon_{0,0}(\mathbf{q},\omega)$. 
In order to include the local field effect, the random-phase-approximation (RPA) calculation is needed to obtain the response function $\epsilon_{\mathrm{mac}} = (\lim_{q \to 0} \epsilon_{0,0}(\mathbf{q},\omega) )^{-1} $. 
The RPA calculation uses the wavefunctions $u$ and the corresponding \textit{k}-space derivatives $\partial u/ \partial k$ generated by the previous IPA calculation and computes the response function $\chi$.

\section{High-throughput workflow}
\subsection{Screening}
As a first step towards building an absorption coefficient database, we consider solids suitable for photolvoltaic applications, specifically semiconductors with an overlapping absorption spectrum with visible light. 
Hence we screened compounds calculated by the Materials Project with three criteria. 
First, an electronic band gap between 0.3 - 3 eV, corresponding to 4100 nm (infra-red) - 413 nm (violet) in photon wavelength; 
Second, to ensure structural stability, the energy above hull was selected to be less than 0.02 eV/atom. This is comparable to the thermal energy $k_B T$ of 25 meV at room temperature which can lead to entropic term that reduce the free energy.
Finally, we limit the number of sites in the unit cell to be less than 10 to ensure a manageable memory footprint for the optical calculations. (Fig.\ref{fig:workflow}) 
This set of filtering results in 1112 candidates, for which the IPA absorption spectra were calculated and integrated into the Materials Project. 

\subsection{Workflow}
\begin{figure}
    \centering
    \includegraphics[scale=0.24]{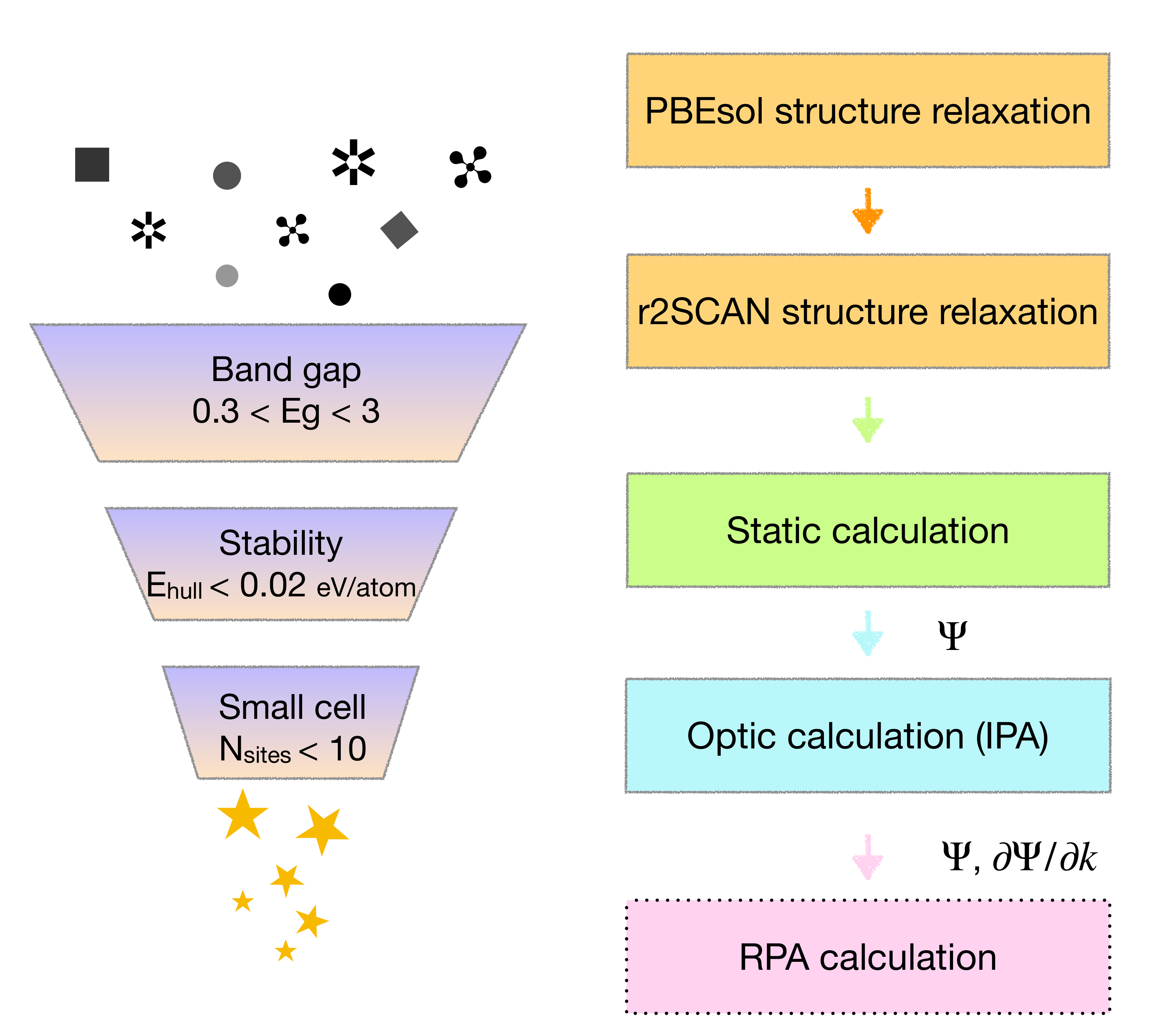}
    \caption{The filtering process and the workflow steps used in this study.}
    \label{fig:workflow}
\end{figure}

We carried out a high-throughput computational workflow on the candidates resulted from the screening step. 
The workflow consists of three parts: 1) structure relaxation 2) static calculation and 3) optic calculation as described in Fig.\ref{fig:workflow}. 
The structural relaxations are carried out using PBEsol functional and an additional relaxation by r2SCAN functional to improve its accuracy.
The static calculation provides a more accurate total energyand the ground state wavefunction. 
The following optical calculation at the IPA level then calculates the dielectric functions, by which the absorption coefficient is derived. 
To examine the importance of including the local field effects, we benchmarked the RPA calculation against IPA for a subset of the materials where experimental results are available. 
The validation is shown in the following section. 

\section{Technical Validation}
\subsection{IPA vs. RPA}

\begin{figure*}
    \centering
    \includegraphics[scale=0.28]{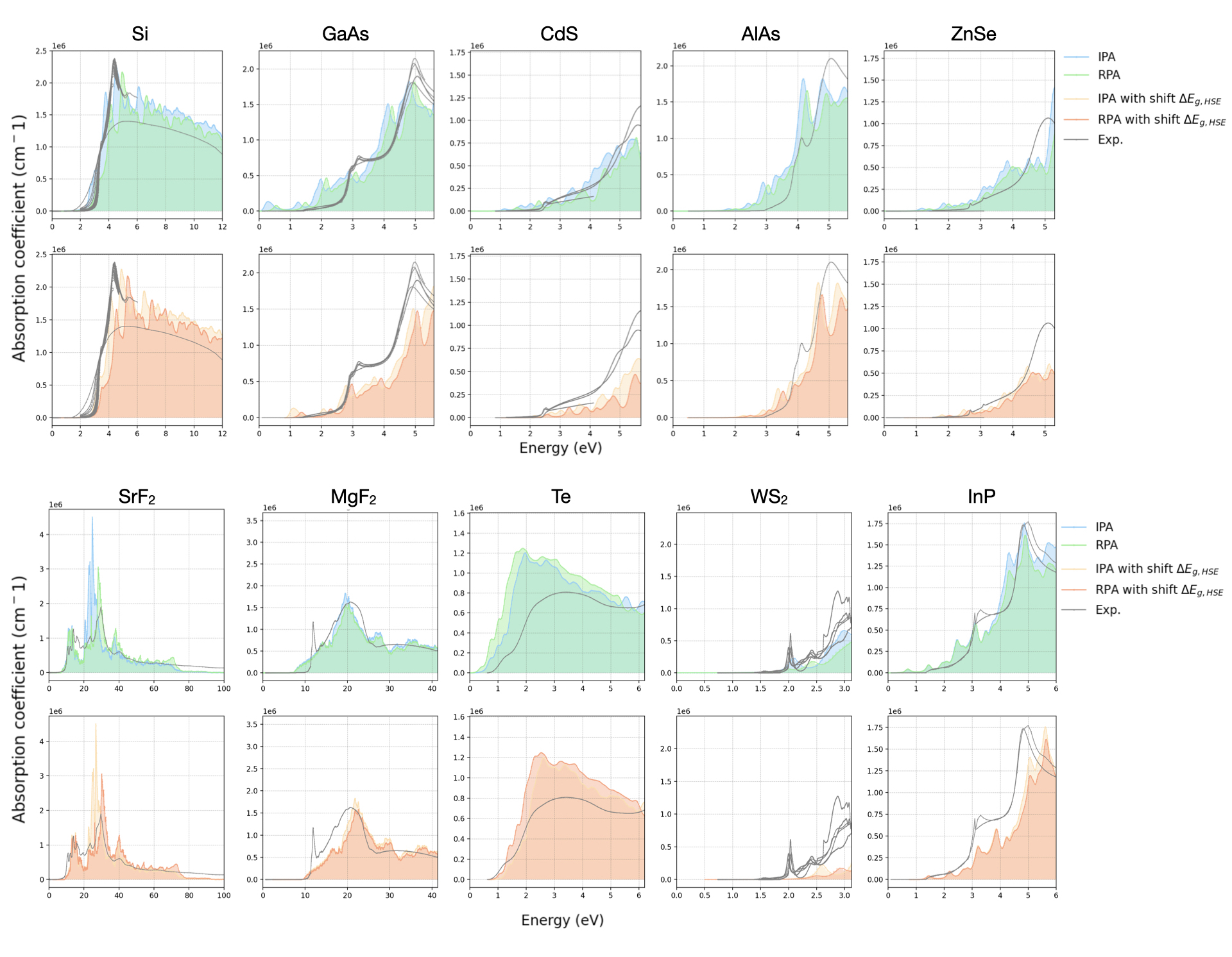}
    \caption{Example comparison between absorption spectrum by IPA, RPA and experiments. Full data sets are included in SI.}
    \label{fig:benchmark}
\end{figure*}

The calculated spectra was benchmarked against experimental results, some of which are shown in Fig \ref{fig:benchmark}. 
Even with the lower-level IPA calculations, the experimental peak shapes and intensities are captured reasonably well. 
Small spectral difference are observed between the IPA and RPA at higher energies, where the local field effects are more prominent. \cite{wiser_dielectric_1963,louie_local-field_1975} 

\begin{figure*}
    \centering
    \includegraphics[scale=0.7]{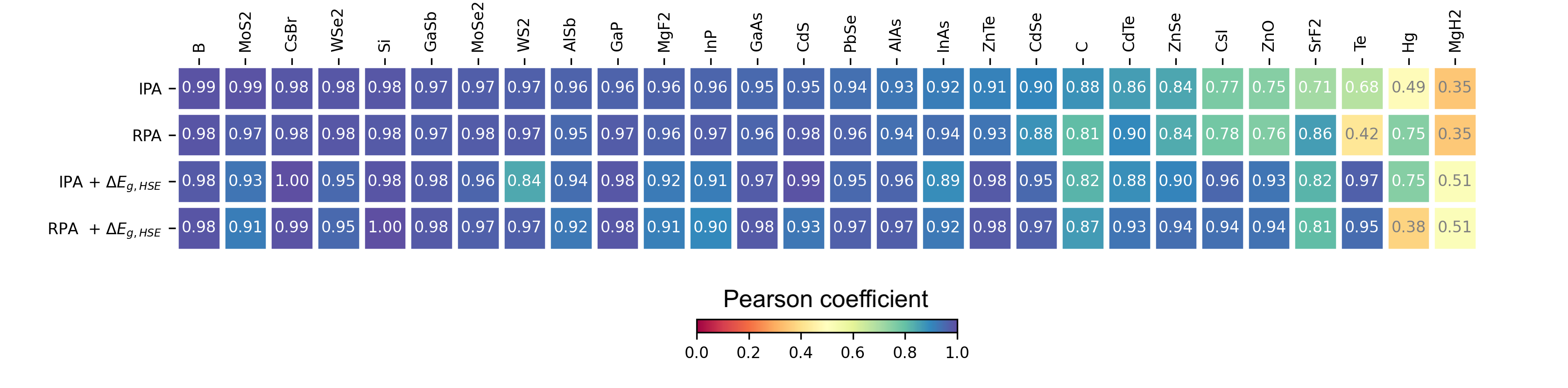}
    \caption{Pearson correlation between simulated spectra and experiments}
    \label{fig:pearson_co}
\end{figure*}

In order to numerically quantify spectral matching, we calculated the Pearson correlation coefficient between the theoretical spectra and the experimental one, using the equation:
\begin{equation}
    P = \frac{\Sigma (x_i - \bar{x})(y_i -\bar{y})}{\sqrt{\Sigma(x_i -\bar{x})^2 \Sigma(y_i -\bar{y})^2}}
\end{equation}
where $x_i$ and $y_i$ are the calculated and experimental absorption coefficients respectively, and $\bar{x}$ and $\bar{y}$ are the average value of the $x_i$ and $y_i$, respectively. 
The Pearson coefficients range between zero and one, and a value closer to 1.0 indicates highly correlated spectra. 
The coefficients are plotted as color map in Fig.\ref{fig:pearson_co}.
For the set of materials with experimental data available, the RPA calculation shows marginal improvement in $P$ in some materials.
In most cases, the differences are negligible, suggesting that the IPA is sufficient to reproduce the spectra. 
Considering the significantly higher cost of the RPA calculation, only the IPA spectrum was calculated for the rest of other compounds ($>$1000). 

We note that for a few materials such as \ce{MgF2} and \ce{WS2}, a sharp peak before the main onset is measured but is absent in our calculations, indicating possible formation of excitons which can only be captured by many-body theories, for example as captured by the Bethe-Salpeter-Equation. 

\subsection{HSE band gap correction}

\begin{figure}
    \centering
    \includegraphics[scale=0.5]{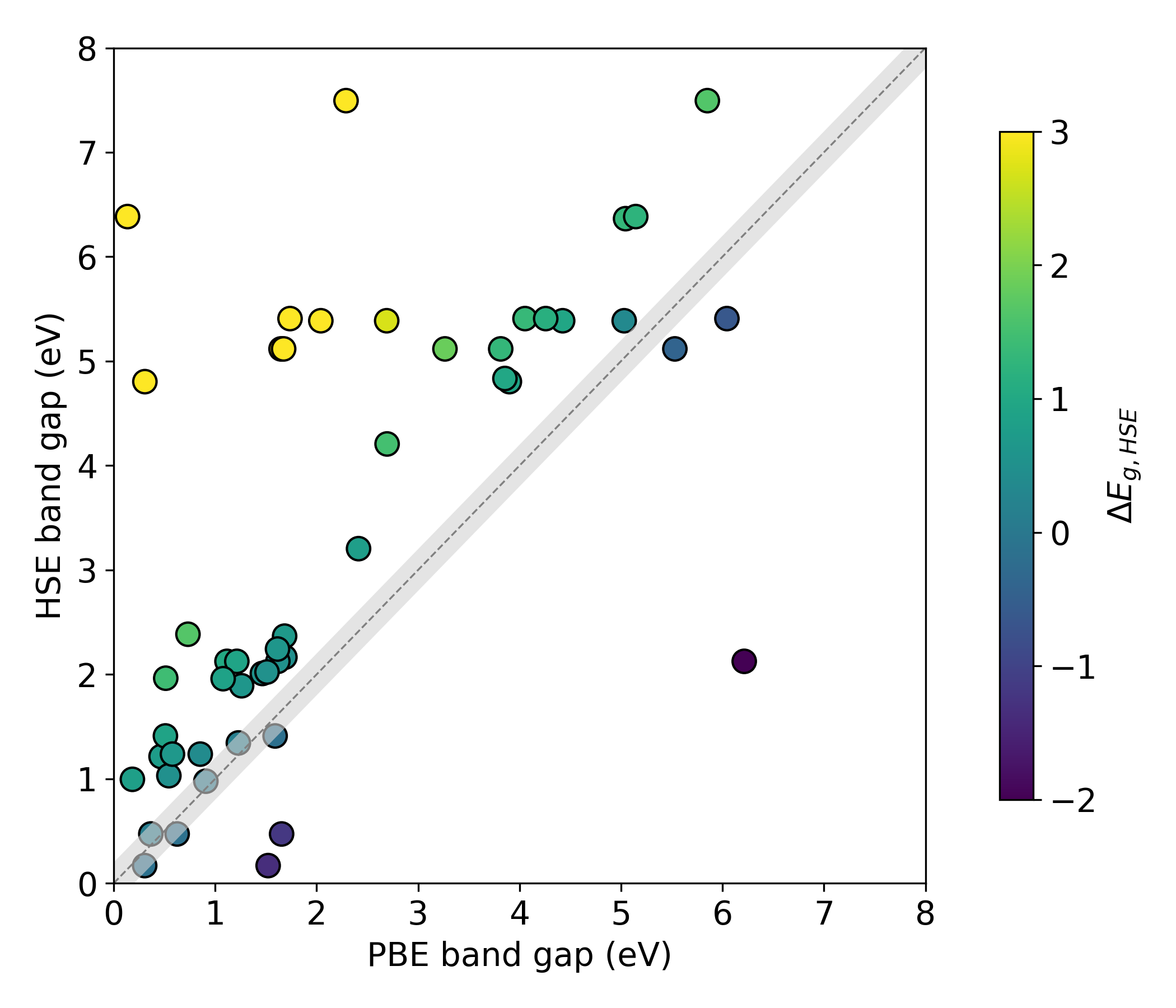}
    \caption{HSE and PBE band gaps for the materials used for benchmark. The color bar represents $\Delta E_{g,\mathrm{,HSE}}$. The gray dashed line corresponds to 1:1 ratio of $E_{g,\mathrm{,HSE}}$/$E_{g,\mathrm{,PBE}}$, and the shaded region represent a $\pm 0.2$ eV window. }
    \label{fig:bandgap}
\end{figure}

For some materials (e.g Si, GaAs, CdS, ZnSe etc.), the onsets of the calculated IPA/RPA absorption are lower than experiments. This is an expected outcome due to the underestimation of bandgaps by the PBE functional.  
To correct the spurious self-interaction in GGA, we used the HSE functional to obtain a more accurate bandgap,\cite{heyd_hybrid_2003-2,heyd_erratum_2006} which includes short-range Hartree-Fock exchange and is known to decrease the error and improve the agreement between the calculated $E_g$ and corresponding experimental values. \cite{crowley_resolution_2016, garza_predicting_2016}
The comparisons in $E_\mathrm{g}$ between HSE and PBE band gaps are shown in Fig.\ref{fig:bandgap}. 
As expected, most of the HSE band gaps are above or close ($<$ 0.2 eV) to the PBE band gaps.
A few outliers include: B ($\Delta E_{g,\mathrm{HSE}}=-1.35$ eV), \ce{Al2O3} ($\Delta E_{g,\mathrm{HSE}}=-0.64$), \ce{H2O} ($\Delta E_{g,\mathrm{HSE}}=-0.41$ eV), \ce{CaSO4} ($\Delta E_{g,\mathrm{HSE}}=-4.09$ eV), \ce{Se} ($\Delta E_{g,\mathrm{HSE}}=-1.18$ eV)

Using the HSE results, the spectra are shifted by $\Delta E_{g,\mathrm{HSE}}$, the energy difference between the PBE band gap and HSE band gap.
This results in a rigid blue shift for most of the spectra, and in many cases better match the experimental data (see Figure \ref{fig:benchmark}). 
A numerical description is reflected in the Pearson correlation for the shifted spectra.Fig.\ref{fig:pearson_co}, where for materials (e.g. CsI, ZnO, Te) with previously lower correlation, this scissor correction improved $P$ significantly. 
This indicates that, to simulate a reasonably accurate absorption spectra, carrying out a PBE-level frequency-dependent dielectric calculation with correction from an HSE band gap is sufficient.

\section{Data Records} 
The calculated spectra are available through the Materials Project website (www.materialsproject.org) and can be downloaded using the REST API.  The results are also available in the form of a JSON file that can be downloaded directly from Materials Project.

\paragraph{Data representations:}
The data for each of the calculated compounds are stored in a list and are provided as a JSON file. For each compound, there are key values, such as ``energy'' and ``density'', that point to the appropriate property (Table 2). Other keys include electronic band gap, imaginary and real part of the dielectric functions, as well as the number of \textit{k}-points.

\begin{table*}[]
\resizebox{\textwidth}{!}{%
\begin{tabular}{@{}lll@{}}
\toprule
json fields             & units   & physical property                                                               \\ \hline
energies                & eV      & Absorption energy corresponding to optic wavelength                             \\
energy\_max              & eV      & The highest energy of the spectrum for this calculation                         \\
optical\_absorption\_co   & cm$^{-1}$ & Absorption coefficient                                                          \\
average\_real\_dielectric & none    & Real part of the frequency-dependent dielectrics, averaged for each tensor      \\
average\_imag\_dielectric & none    & Imaginary part of the frequency-dependent dielectrics, averaged for each tensor \\
bandgap                 & eV      & The electronic band gap                                                         \\
nkpoints                & none    & The number of kpoints used in the optic calculation                             \\ 
\hline
\end{tabular}%
}
\label{tab:json}
\caption{The data fields of a downloadable absorption spectrum from MP.}
\end{table*}

\section{Usage Note} 
We present a database of calculated optical absorption spectra for 1,116 compounds. The data should be of broad interest to materials applications relating to electronic structure, particularly optoelectronic and other electronic properties. 
For example, we expect this database to be used in the understanding of optical absorption and in the search for new solar absorbers with unique and tailored properties.
The above use cases are facilitated by the Materials Project website interface which allows users to search for materials with associated metrics, such as stability, band gap and/or density. 
With this new dataset and underlying data and software infrastructure, users will now be able to request calculated optical absorption coefficients and frequency-dependent dielectric functions.  Furthermore, these capabilities open up opportunities in data-driven endeavors, such as the application of machine learning techniques to identify structural and chemical features that are key to highly absorbing materials, hence extrapolating the data beyond its current coverage and accelerating the discovery of novel optoelectronic materials.

\section{Acknowledgement}
This research used resources of the National Energy Research Scientific Computing Center (NERSC), a U.S. Department of Energy Office of Science User Facility located at Lawrence Berkeley National Laboratory, operated under Contract No. DE-AC02-05CH11231. We thank Rachel Woods-Robinson for helpful discussions. The work is supported by the US Department of Energy, Office of Science, Office of Basic Energy Sciences, Materials Sciences and Engineering Division under contract no. DE-AC02-05-CH11231 (Materials Project program KC23MP). 

\bibliography{absorption_wf}

\end{document}